\documentclass[aps,prd,twocolumn,10pt]{revtex4-1}
\usepackage{amssymb}
\usepackage{graphicx}
\usepackage{amsmath}
\usepackage{hyperref}
\usepackage{subfigure}
\usepackage{multirow}
\usepackage{setspace}
\usepackage{verbatim}
\usepackage{float}
\usepackage{color}
\usepackage{ulem}
\usepackage[utf8]{inputenc}

\begin{document}

\title{No-go guide for the Hubble tension : Late-time solutions}

\author{Rong-Gen Cai$^{1,2,3}$}
\email{cairg@itp.ac.cn}

\author{Zong-Kuan Guo$^{1,2,3}$}
\email{guozk@itp.ac.cn}

\author{Shao-Jiang Wang$^{1}$}
\email{schwang@itp.ac.cn (corresponding author)}

\author{Wang-Wei Yu$^{1,3}$}
\email{yuwangwei@mail.itp.ac.cn (corresponding author)}

\author{Yong Zhou$^{1}$}
\email{zhouyong@itp.ac.cn}

\affiliation{$^1$CAS Key Laboratory of Theoretical Physics, Institute of Theoretical Physics, Chinese Academy of Sciences, Beijing 100190, China}
\affiliation{$^2$School of Fundamental Physics and Mathematical Sciences, Hangzhou Institute for Advanced Study (HIAS), University of Chinese Academy of Sciences (UCAS), Hangzhou 310024, China}
\affiliation{$^3$School of Physical Sciences, University of Chinese Academy of Sciences (UCAS), Beijing 100049, China}

\begin{abstract}
The Hubble tension, if not caused by any systematics, could be relieved or even resolved from modifying either the early-time or late-time Universe. The early-time modifications are usually in tension with either galaxy clustering or galaxy lensing constraints. The late-time modifications are also in conflict with the constraint from the inverse distance ladder, which, however, is weakened by the dependence on a sound-horizon prior and some particular approximation for the late-time expansion history. To achieve a more general no-go argument for the late-time scenarios, we propose to use a global  parametrizationbased on the cosmic age (PAge) to consistently use the cosmic chronometers data beyond the Taylor expansion domain and without the input of a sound-horizon prior. Both the early-time and late-time scenarios are therefore largely ruled out, indicating the possible ways out of the Hubble tension from either exotic modifications of our concordance Universe or some unaccounted systematics.
\end{abstract}
\maketitle

\section{Introduction} 

The extrapolation from globally fitting the $\Lambda$-cold-dark-matter ($\Lambda$CDM) model to the cosmic microwave background (CMB) data \cite{Henning:2017nuy,Aghanim:2018eyx,Aiola:2020azj} renders the Hubble constant $H_0=67.27\pm0.60\,\mathrm{km\,s^{-1}\,Mpc^{-1}}$ to an unprecedented accuracy \cite{Aghanim:2018eyx}. On the other hand, the combined big bang nucleosynthesis (BBN) + baryon acoustic oscillation (BAO) constraint  \cite{Addison:2013haa,Aubourg:2014yra,Addison:2017fdm,DES:2017txv,Cuceu:2019for,Schoneberg:2019wmt,Philcox:2020vvt,DAmico:2020kxu,DES:2021wwk} is independent of the CMB data, yet still shares a similar $H_0$ value as inferred by CMB data if $\Lambda$CDM is assumed throughout the early Universe. Furthermore, the consistency of $\Lambda$CDM with respect to the early-Universe observations is also manifested in the consistency tests of the early integrated Sachs-Wolfe effect \cite{Vagnozzi:2021gjh} and the sound horizon measured at the matter-radiation equality, recombination, and the end of the drag epoch \cite{Philcox:2020xbv,Lin:2021sfs} . Apart from some anomalies that arise in the high-$\ell$ \cite{Aylor:2017haa,Henning:2017nuy,Knox:2019rjx}  or EE-polarized \cite{SPT-3G:2021eoc,Addison:2021amj} CMB data, the early-Universe observations could, at the very least, achieve a consensus on the Hubble constant $H_0\lesssim70\,\mathrm{km\,s^{-1}\,Mpc^{-1}}$.

However, the Hubble constants inferred by $\Lambda$CDM  from early-Universe observations are systematically lower than those from local measurements either depending on or independent of the distance ladders. The local measurements with distance ladders heavily rely on the calibrations from Cepheid \cite{Riess:2016jrr,Riess:2018byc,Riess:2018uxu,Riess:2019cxk,Riess:2020fzl}, tip of the red giant branch (TRGB) \cite{Freedman:2019jwv,Yuan:2019npk,Freedman:2020dne,Soltis:2020gpl,Freedman:2021ahq}, or Miras \cite{Huang:2019yhh} for connecting the local geometric measurements with the type Ia supernovae (SNe) in the Hubble flow. Nevertheless, the most recent measurement $H_0=69.8\pm2.2\,\mathrm{km\,s^{-1}\,Mpc^{-1}}$ from TRGB calibration \cite{Freedman:2021ahq} is lower than the most recent Cepheid calibration result $H_0=73.2\pm1.3\,\mathrm{km\,s^{-1}\,Mpc^{-1}}$  \cite{Riess:2020fzl}, which might be affected by the choice of Cepheid color-luminosity calibration method \cite{Mortsell:2021nzg} and other sources of uncertainty in the supernova distance ladder \cite{Mortsell:2021tcx}.   On the other hand, the local measurements independent of distance ladders vary largely among megamaser \cite{Huang:2019yhh,Pesce:2020xfe}, surface brightness fluctuations  \cite{Khetan:2020hmh,Blakeslee:2021rqi}, baryonic Tully–Fisher relation \cite{Kourkchi:2020iyz,Schombert:2020pxm}, gravitational-wave standard sirens \cite{Abbott:2017xzu,LIGOScientific:2019zcs,Mukherjee:2019qmm,Wang:2020vgr,Wang:2020dkc}, strong lensing time delay (SLTD) \cite{Wong:2019kwg,Shajib:2019toy,Birrer:2020tax}, parallax measurement of quasar 3C 273 \cite{Wang:2019gaq}, and extragalactic background light $\gamma$-ray  attenuation \cite{Dominguez:2019jqc}. At the very least, the local measurements from late-Universe observations seem to agree on $H_0\gtrsim70\,\mathrm{km\,s^{-1}\,Mpc^{-1}}$.

The $\sim4\sigma$ Hubble tension \cite{Bernal:2016gxb,Verde:2019ivm,Knox:2019rjx,Riess:2020sih,DiValentino:2020zio} only arises when confronting the Planck measurement \cite{Aghanim:2018eyx} with the Cepheid measurement  \cite{Riess:2020fzl}, the most precise measurement from each camp. However, the rest of comparisons drawn from early-time and late-time observations are insufficient to claim a significant tension but with a rough compatibility around $H_0\simeq70\,\mathrm{km\,s^{-1}\,Mpc^{-1}}$ \cite{Freedman:2021ahq}. In perspective of future developments, there are two possibilities to pursue: 1) If the Hubble tension is not real, then it should be feasible to show the consistency of $\Lambda$CDM with late-time data independent of CMB and local $H_0$ measurements. 2) If the Hubble tension is real, then it is necessary to narrow down the possible models \cite{DiValentino:2020zio,DiValentino:2021izs}  either from early-time or late-time scenarios:

(i) For early-time solutions, one can modify either the expansion history or recombination history. The early-time expansion history could be altered by some temporary energy injection around the matter-radiation equality, for example, dark radiation (DR) 
and early dark energy (EDE) \cite{Karwal:2016vyq,Poulin:2018dzj,Poulin:2018cxd,Agrawal:2019lmo,Lin:2019qug,Berghaus:2019cls,Sakstein:2019fmf,Smith:2019ihp,Niedermann:2019olb,Ye:2020btb,Gonzalez:2020fdy}.
The free-streaming DR is strongly constrained by BAO+BBN \cite{Schoneberg:2019wmt} before the BBN epoch and disfavored by the absence of the neutrino free-streaming phase shift in CMB  \cite{Knox:2019rjx}.  The non-free-streaming DR, for example, strongly self-interacting neutrinos \cite{Kreisch:2019yzn}, is also disfavored by the high-$\ell$ polarisation CMB data \cite{Das:2020xke,RoyChoudhury:2020dmd}. The EDE models also deteriorate the $S_8$ tension \cite{Hill:2020osr,Ivanov:2020ril,DAmico:2020ods} (see, however, Refs. \cite{Chudaykin:2020acu,Chudaykin:2020igl,Niedermann:2020qbw,Murgia:2020ryi,Smith:2020rxx}) and BBN constraint \cite{Seto:2021xua}. On the other hand, changing the recombination history \cite{Jedamzik:2020krr,Chiang:2018xpn,Hart:2019dxi,Sekiguchi:2020teg} (see also Ref. \cite{Liu:2019awo}) via primordial magnetic fields \cite{Jedamzik:2020krr} found no evidence for the required baryon clumping  \cite{Thiele:2021okz}. In summary, a general no-go argument \cite{Jedamzik:2020zmd} could be put forward that, for early-time solutions that solely reduce the cosmic sound horizon, models with lower values of $\Omega_m h^2$ are in tension with galaxy clustering data \cite{BOSS:2016wmc}, while models with higher values of $\Omega_m h^2$ are in tension with galaxy weak lensing data \cite{DES:2017myr,KiDS:2020suj}. This therefore largely rules out early-time solutions.  

(ii) For various late-time solutions, the $H_0$ constraints from the SNe data with their absolute magnitude calibrated by Cepheid variables are quasi-model-independent \cite{Dhawan:2020xmp}. However, the situation changes when including BAO data. This is the usual no-go argument for the late-time solutions using the inverse distance ladder \cite{Cuesta:2014asa,Heavens:2014rja,Aubourg:2014yra,Verde:2016ccp,Alam:2016hwk,Verde:2016wmz,Macaulay:2018fxi,Feeney:2018mkj,eBOSS:2020yzd}, which combines BAO+SNe with a CMB prior on the sound horizon $r_s$ \cite{Vonlanthen:2010cd,Audren:2012wb,Audren:2013nwa,Cuesta:2014asa,Verde:2016ccp,Bernal:2016gxb,Verde:2016wmz,Aylor:2018drw} since BAO can only constrain the combinations $H(z)r_s$ and $D_A(z)/r_s$. Note that $r_s$ is mainly determined by the early-Universe evolution (thus independent of late-time evolution) and therefore unharmful to be used to discriminate the late-time models.  To implement the inverse distance ladder \cite{Lemos:2018smw}, one first assumes a Planck's prior on $r_s\simeq147\,\mathrm{Mpc}$ and some phenomenological parametrizationfor $H(z)$ at late times and then fits the combined BAO+SNe data with an astrophysical determination on the SNe Ia absolute magnitude $M_B$ \cite{Riess:2020fzl}, leading to a strong constraint on the late-time Universe to be barely deviated from $\Lambda$CDM within $0.01\lesssim z\lesssim 1$. Although a sudden phantom transition below $z\lesssim0.01$ seems to raise the local $H_0$ value while still maintaining the phenomenological success of $\Lambda$CDM above $z\gtrsim0.01$ \cite{Mortonson:2009qq}, the price to pay is to deviate $M_B$   fitted by CMB+BAO+SNe significantly from the one used to derive a locally higher $H_0$ \cite{Benevento:2020fev,Camarena:2021jlr,Efstathiou:2021ocp}. This therefore largely rules out phantomlike dark energy models.

A more general no-go argument \cite{Zhang:2020uan} without a CMB prior on $r_s$ (thus also independent of the early-time cosmology) could be made by combining BAO+SNe with observational $H(z)$ data (OHD) for some late-time $H(z)$ parameterizations from Taylor expansions in $z$ or $(1-a)$ \cite{Cattoen:2007sk}. However, the Taylor expansions of $H(z)$ in $z$ or $(1-a)$ even to the fourth order still fail to cover the OHD redshift with the modest accuracy even for the $\Lambda$CDM case. We therefore propose in this paper to use a global parametrizationbased on the cosmic age (PAge) \cite{Huang:2020mub,Luo:2020ufj} that not only reproduces $\Lambda$CDM up to high redshift with high accuracy but also covers a large class of late-time models in a wide redshift range with a high accuracy. Furthermore, it is logically more consistent to use PAge for OHD from the cosmic chronometer (CC), and the cosmic age was recently found to play an important role in the Hubble tension \cite{Jimenez:2019onw,Bernal:2021yli,Vagnozzi:2021tjv,Boylan-Kolchin:2021fvy}. Whether the Hubble tension turns out to be real or not, our work could serve as either a no-go guide beyond or a consistency test for the $\Lambda$CDM model, respectively.

\section{Model}

\begin{figure}
\centering
\includegraphics[width=0.48\textwidth]{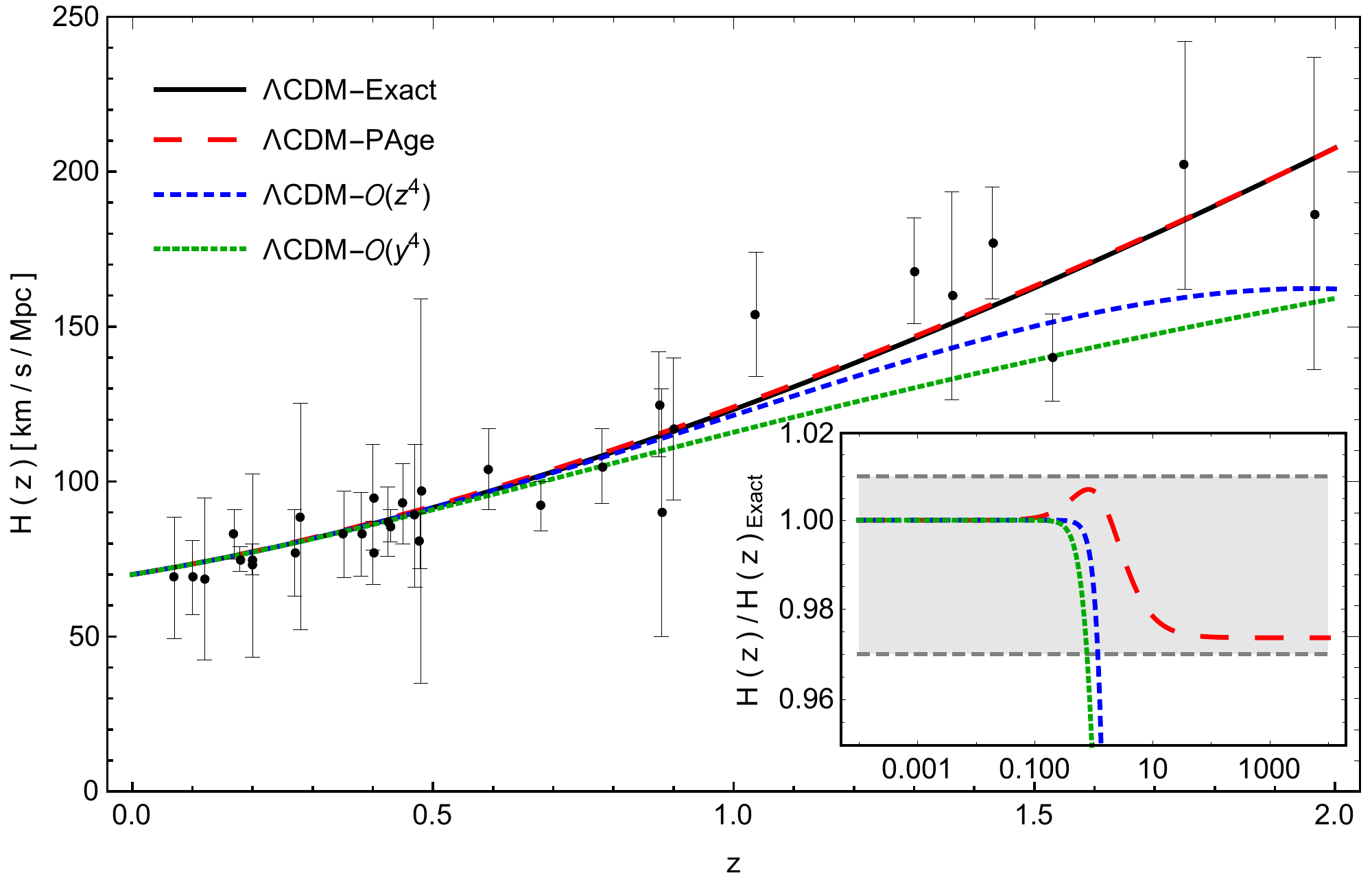}
\includegraphics[width=0.48\textwidth]{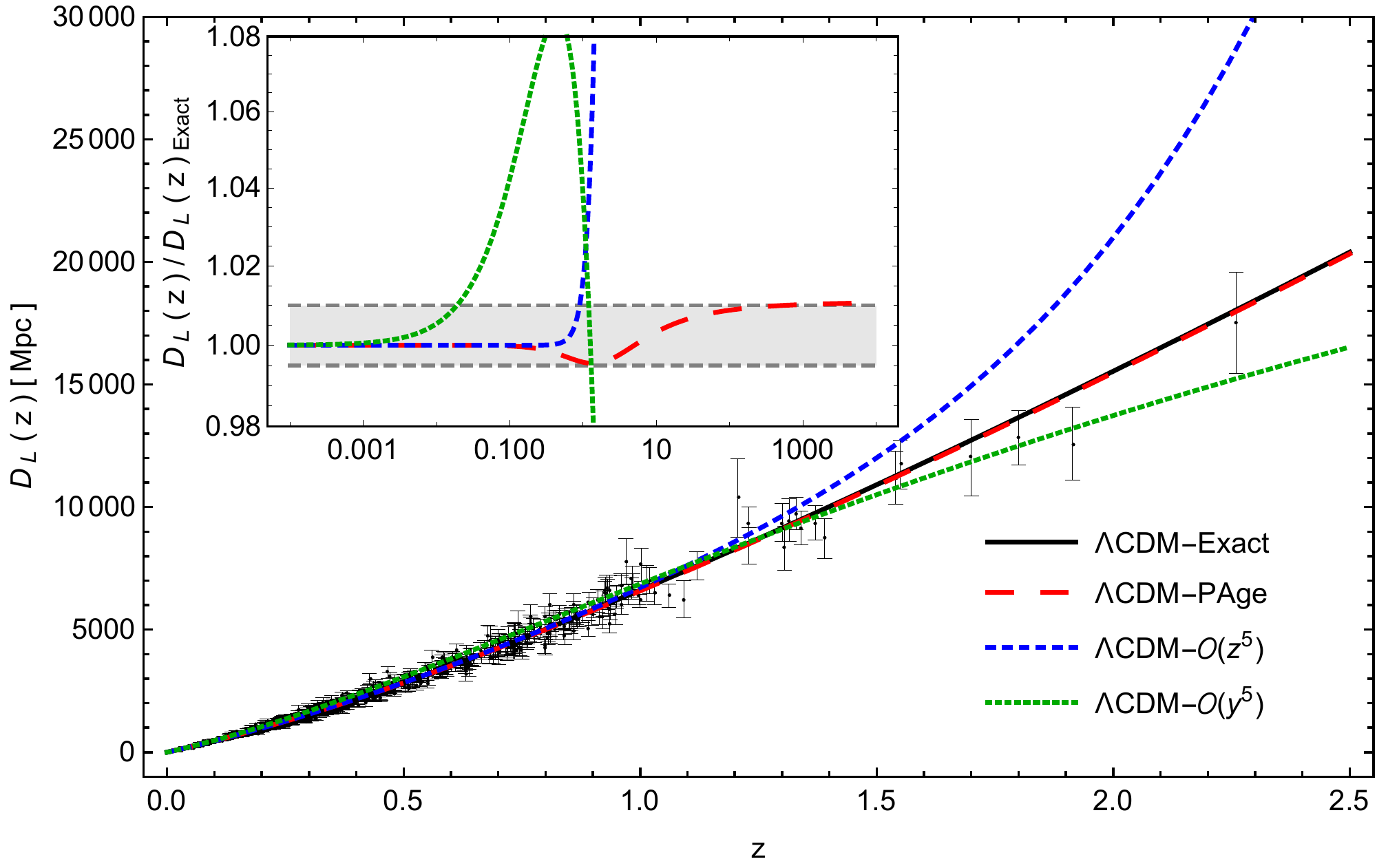}\\
\caption{The comparison between the PAge model (red dashed) and Taylor expansions in redshift  $z$ (blue dashed) and $y$-redshift (green dashed) compared to the exact expression from $\Lambda$CDM (black) for the Hubble expansion rate $H(z)$ and luminosity distance $D_L(z)$ with fiducial cosmology  $\Omega_m=0.3$, $H_0=70\,\mathrm{km\,s^{-1}\,Mpc^{-1}}$. The $H(z)$ data and Pantheon data (converted into luminosity distances with fiducial value $M_B=-19.34$) are shown for illustration. The relative errors are shown in the inserts for a larger redshift range up to $z\sim10^4$. }\label{fig:HzDL}
\end{figure}

The usual model-independent parametrizationfor the late-time expansion history adopts a Taylor expansion \cite{Cattoen:2007sk} either in redshift $z$ \cite{Visser:2003vq,Zhang:2016urt} or in $y$-redshift $y\equiv1-a=z/(1+z)$  \cite{Capozziello:2011tj} as shown in Appendix A of the Supplemental Material \cite{footnote} for the dimensionless Hubble expansion rate $E=H/H_0$ and dimensionless luminosity distance $d_L=D_L/(c/H_0)$. Although the Taylor expansion in  $y$-redshift slightly improves the convergence of the Taylor expansion in redshift $z$, both of them still deviate significantly from the exact formula even for the $\Lambda$CDM case as shown in Fig. \ref{fig:HzDL} with blue and green dashed lines. Introducing more terms with higher orders in $z$ or $y$ could certainly improve the convergence behavior but also weaken the constraining power of data fitting due to the presence of more nuisance parameters.

PAge is introduced as a global approximation of the cosmic expansion history \cite{Huang:2020mub}. Assuming our Universe is dominated by the matter component at high redshift $z\gg1$ and ignoring the radiation component and the very short period of radiation dominance before matter dominance, one could approximate the product of the Hubble expansion rate $H$ and the cosmological time $t$  as a quadratic function of $t$, namely,
\begin{align}\label{eq:PAge}
\frac{H}{H_0}=1+\frac23\left(1-\eta\frac{H_0t}{p_\mathrm{age}}\right)\left(\frac{1}{H_0t}-\frac{1}{p_\mathrm{age}}\right),
\end{align}
where the parameter $\eta$ could be evaluated as
\begin{align}\label{eq:eta}
\eta=1-\frac32p_\mathrm{age}^2(1+q_0)
\end{align}
by taking a time derivative of $H$ in \eqref{eq:PAge} followed by a replacement of the deceleration parameter $q(t)=-\ddot{a}a/\dot{a}^2$. $p_\mathrm{age}\equiv H_0t_0$ is the product of $H_0\equiv100h\,\mathrm{km\,s^{-1}\,Mpc^{-1}}$ and the current age of Universe $t_0$. 

For $\Lambda$CDM with late-time parametrization $H(a)=H_0\sqrt{\Omega_ma^{-3}+(1-\Omega_m)}$, one has $q_0=-1+\frac32\Omega_m$, and the current age of our Universe reads
\begin{align}
t_0=\int_0^1\frac{\mathrm{d}a}{aH(a)}=\frac{9.77788\mathrm{Gyr}}{3h\sqrt{1-\Omega_m}}\ln\frac{1+\sqrt{1-\Omega_m}}{1-\sqrt{1-\Omega_m}}.
\end{align}
Therefore, $\eta=0.3726$ and $p_\mathrm{age}=0.9641$ for fiducial $\Lambda$CDM with $\Omega_m=0.3$ and $H_0=70\,\mathrm{km\,s^{-1}\,Mpc^{-1}}$. The corresponding  $H(z)$ and $D_L(z)$ are shown in Fig. \ref{fig:HzDL} with red dashed lines, which differ from the exact $\Lambda$CDM expressions below  $3\%$ and $1\%$, respectively, over the whole redshift range up to $z\sim10^4$ as shown in the inserts. 

For models beyond $\Lambda$CDM, both parameters $\eta$ and $p_\mathrm{age}$ should be treated as free parameters and the only two free parameters in $H/H_0$ of  \eqref{eq:PAge}. To see this, we can directly solve \eqref{eq:PAge} for the combination $H_0t$ after replacing $H$ with $-\mathrm{d}z/\mathrm{d}t/(1+z)$, namely,
\begin{align}
1+z=\left(\frac{p_\mathrm{age}}{H_0t}\right)^\frac23e^{\frac13\left(1-\frac{H_0t}{p_\mathrm{age}}\right)\left(3p_\mathrm{age}+\eta\frac{H_0t}{p_\mathrm{age}}-\eta-2\right)},
\end{align}
then $H_0t$ in \eqref{eq:PAge} is a function of $z$, leaving only two free parameters $\eta$ and $p_\mathrm{age}$ in $H/H_0$ of  \eqref{eq:PAge}. For a specific physical model, $\eta$ and $p_\mathrm{age}$  could be expressed by the model parameters. Mapping a specific  model in the PAge parameter space requires matching $q(t)$ at some characteristic time, for example, at redshift zero, as done in Table 1 of Ref. \cite{Luo:2020ufj} for a large class of  illustrative models, where the relative error for the PAge representation of the $ow_\mathrm{CPL}\mathrm{CDM}$ model \cite{Chevallier:2000qy,Linder:2002et} is less than $1\%$ over $0<z<2.5$. See Appendix B in the supplemental material \cite{footnote} for more details on model matching. Note that the focus of Ref. \cite{Huang:2020mub} for proposing the PAge parametrizationis to reconfirm the late-time acceleration from a lower bound on $t_0>12$ Gyr, where SNe data with a $H_0$ prior and a CMB distance prior are used for data analysis. This is totally different from the purpose of this paper and the data analysis strategy as presented below.

\section{Data analysis}

\begin{table*}
\caption{The cosmological constraints from fitting the datasets SNe+BAO+OHD(BC03) and SNe+BAO+OHD(MS11) to the $\Lambda$CDM and PAge models with free parameters $\{\Omega_m, M_B, H_0, r_d\}$ and $\{\eta, p_\mathrm{age}, M_B, H_0, r_d\}$, respectively. }\label{tab:constraint}
\begin{tabular}{c|c|c|c|c|c}
\hline
\hline
\multirow{2}*{Parameter} 
& \multirow{2}*{Prior range} 
& \multicolumn{2}{|c}{BC03} 
& \multicolumn{2}{|c}{MS11} \\
\cline{3-6} &  & $\Lambda$CDM & PAge model & $\Lambda$CDM  & PAge model \\
\hline
$\Omega_m$           & $0.15\sim0.5$   & $0.288\pm0.011$                    & $-$                                              & $0.282\pm0.011$                      & $-$\\
$\eta$                       & $-2\sim2$        & $-$                                            & $0.334_{-0.057}^{+0.067}$       &  $-$                                          & $0.341_{-0.063}^{+0.059}$ \\
$p_\mathrm{age}$   & $0.15\sim2.0$  & $-$                                            & $0.975_{-0.011}^{+0.012}$         &  $-$                                          & $0.978_{-0.009}^{+0.010}$ \\
$M_B$                      &  $-20\sim-19$  & $-19.374\pm0.047$                 & $-19.379_{-0.052}^{+0.051}$    & $-19.309_{-0.053}^{+0.059}$ & $-19.322\pm0.063$ \\
$H_0$                       &  $60\sim80$    & $69.389_{-1.474}^{+1.547}$    & $68.958_{-1.826}^{+1.779}$    & $71.591_{-1.780}^{+1.950}$   & $70.799_{-2.118}^{+2.220}$ \\
$r_d$                         &  $120\sim160$ & $146.563_{-2.894}^{+3.293}$ & $146.466_{-3.302}^{+3.448}$ & $142.573_{-3.820}^{+3.754}$  & $142.885_{-4.062}^{+4.287}$ \\
\hline
$\chi_\mathrm{min}^2/\mathrm{d.o.f}$ & $-$            & $0.9724$                                   & $0.9708$                                   & $0.9803$                                   & $0.9789$ \\
\hline
\hline
\end{tabular}
\end{table*}

\begin{figure*}
\centering
\includegraphics[width=0.48\textwidth]{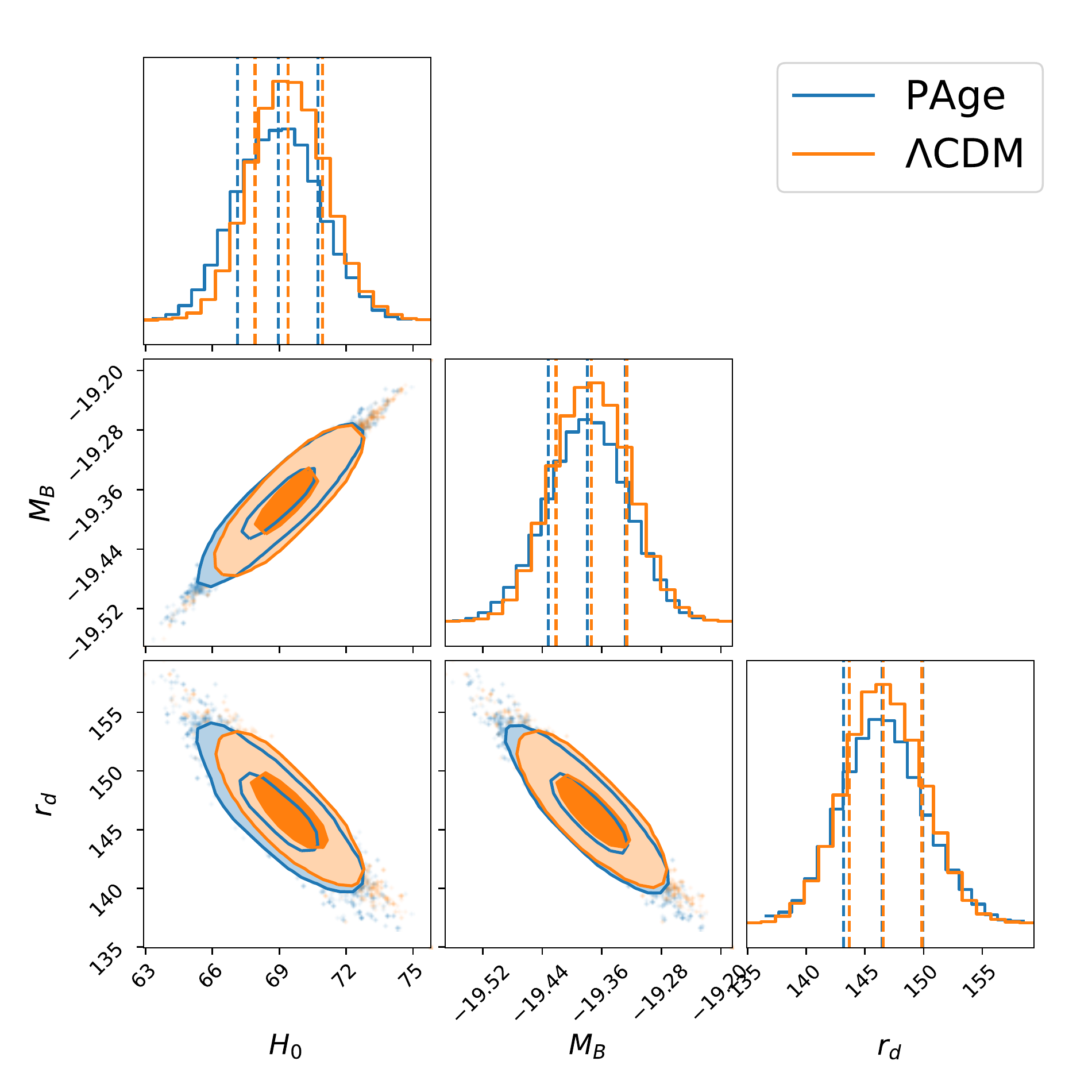}
\includegraphics[width=0.48\textwidth]{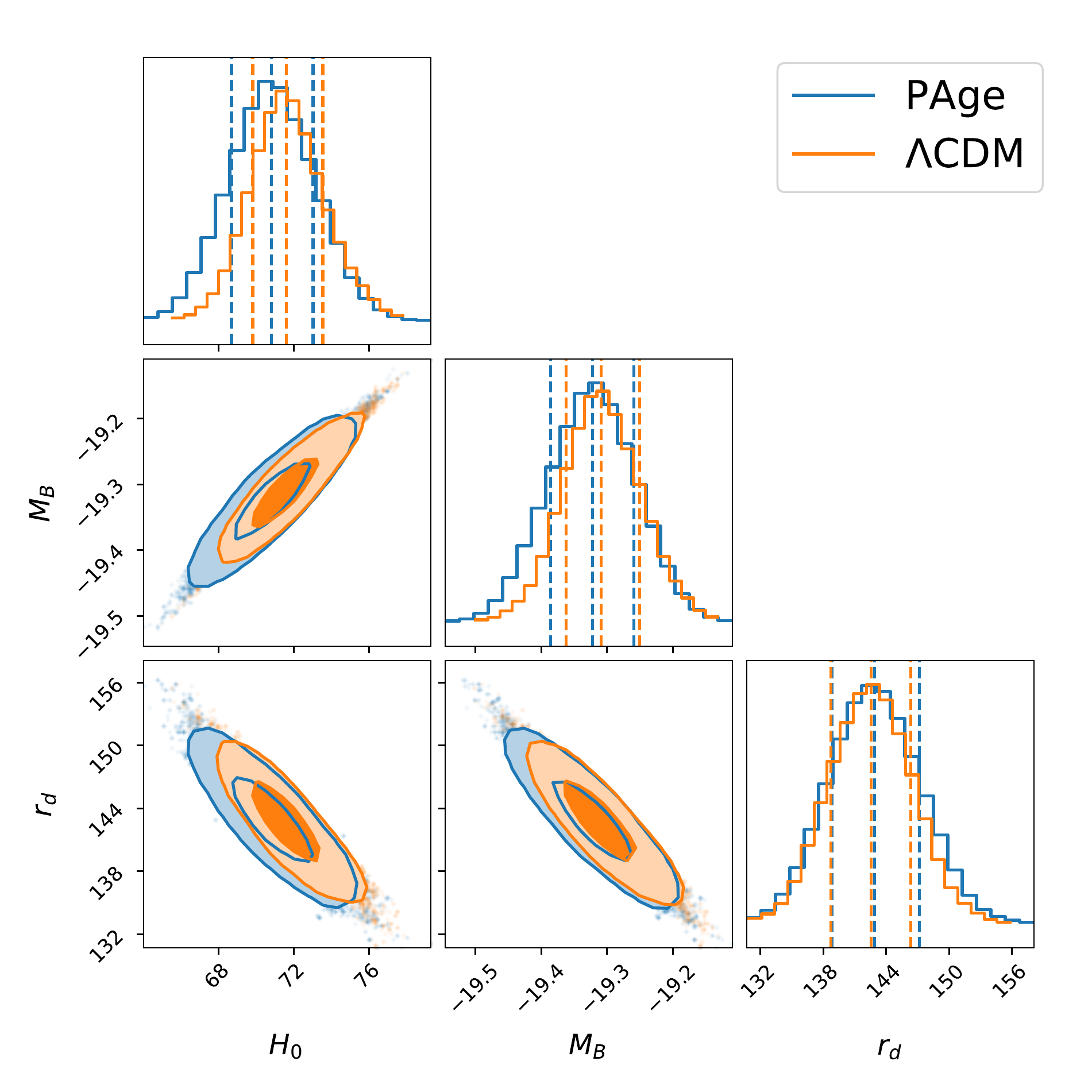}\\
\caption{$1\sigma$ and $2\sigma$ constraints from fitting the datasets SNe+BAO+OHD(BC03) (left panel) and SNe+BAO+OHD(MS11) (right panel) to the $\Lambda$CDM and PAge models with free parameters $\{\Omega_m, M_B, H_0, r_d\}$ and $\{\eta, p_\mathrm{age}, M_B, H_0, r_d\}$, respectively. The fitting results involving with $\eta$, $p_\mathrm{age}$, and $\Omega_m$ are not shown here.}\label{fig:MCMC}
\end{figure*}

The data we use includes SNe Ia (standard candle), BAO (standard ruler), and OHD (standard clock), which is independent of either local $H_0$ measurements or the early-Universe observations like CMB and BBN.

For SNe Ia data, we use the Pantheon sample \cite{Scolnic:2017caz} containing 1048 SNe Ia within $0.01<z<2.3$. The SNe data directly measure the apparent magnitude $m_B(z)$, which could be computed theoretically from a model by
\begin{align}
m_B(z)=M_B+5\lg\frac{D_L(z)}{10\,\mathrm{pc}}=a_B+5\lg d_L(z).
\end{align}
For a given model with dimensionless Hubble expansion rate $E=H/H_0$, the dimensionless luminosity distance is known as
\begin{align}
d_L(z)\equiv\frac{D_L(z)}{c/H_0}=(1+z)\int_0^z\frac{\mathrm{d}z'}{E(z')}.
\end{align}
What the SNe magnitude-redshift relation actually constrains is  $a_B\equiv M_B+42.3841-5\lg h$. With a $M_B$ prior from local distance ladders, one infers the value of $H_0$. However, as pointed out in Refs.  \cite{Benevento:2020fev,Camarena:2021jlr,Efstathiou:2021ocp}, this inferred $H_0$ might not be consistent with the constraint on $H_0$ if one sets both $H_0$ and $M_B$ free in the inverse distance ladder. In other words, if one adopts the SH0ES's prior on $H_0$ for a certain model, the inferred $M_B$ from $a_B$ might not be consistent with the constraint on $M_B$ if both $H_0$ and $M_B$ are free in the inverse distance ladder. As a result, both $H_0$ and $M_B$ will be regarded as free parameters  in the inverse distance ladder.

For BAO data, we use the state-of-the-art datasets \cite{Beutler:2011hx,Ross:2014qpa,BOSS:2016zkm,Ata:2017dya,deSainteAgathe:2019voe,Blomqvist:2019rah,Bautista:2020ahg,Gil-Marin:2020bct,deMattia:2020fkb,Tamone:2020qrl,Neveux:2020voa,Hou:2020rse,duMasdesBourboux:2020pck,DES:2021esc} as listed in  Appendix C of the Supplemental Material \cite{footnote}. The BAO measurements are summarized at some effective redshifts $z_\mathrm{eff}$ for the BAO feature in both line-of-sight and transverse directions. Along the line-of-sight direction, BAO directly measures $D_H(z)/r_d$ with respect to some fiducial cosmology, where the Hubble distance is defined as
\begin{align}
D_H(z)=\frac{c}{H(z)}.
\end{align}
Along the transverse direction, BAO directly measures $D_M(z)/r_d$ or $D_A(z)/r_d$ with respect to the same fiducial cosmology, where the (comoving) angular diameter distances are defined via
\begin{align}
D_M(z)=\frac{D_L(z)}{1+z}=(1+z)D_A(z).
\end{align}
For historical reason, BAO measurements could also be summarized by $D_V(z)/r_d$, where the spherically averaged distance is defined as
\begin{align}
D_V(z)=\left[zD_M(z)^2D_H(z)\right]^{1/3}. 
\end{align}
To detach the model dependence on the early-Universe cosmology and observations, the sound horizon at drag epoch $r_d$ will be treated as a free parameter.

For OHD from the differential age method \cite{Jimenez:2001gg}, the Hubble parameter could be directly measured by
\begin{align}
H(z)=-\frac{1}{1+z}\frac{\mathrm{d}z}{\mathrm{d}t}
\end{align}
from the age difference $\Delta t$ between two passively evolving galaxies that formed at the same time but are separated by a small redshift interval $\Delta z$. This method is independent of any cosmological models but the age estimation on the evolutionary stellar population synthesis (EPS) models. We use OHD \cite{Moresco:2016mzx,Moresco:2012jh,Zhang:2012mp,Ratsimbazafy:2017vga,Stern:2009ep,Moresco:2015cya} as listed in Appendix C of the Supplemental Material \cite{footnote} from two different EPS models: Bruzual and Charlot (2003) \cite{Bruzual:2003tq} (BC03 hereafter) and Maraston and  Str\"{o}mb\"{a}ck (2011) \cite{Maraston:2011sq} (MS11 hereafter). Note that the OHD points at $z=1.363$ and $z=1.965$ from \cite{Moresco:2015cya} have adopted both EPS models of BC03 and MS11, which will not be included in the results presented below. Nevertheless, we have checked that the naive inclusion of these two OHD points in both datasets has little impact on our results and conclusions.

Fitting above SNe+BAO+OHD(BC03/MS11) to the PAge model with $\{\eta, p_\mathrm{age}, M_B, H_0, r_d\}$ as the free parameters with flat priors as listed in Table \ref{tab:constraint}, we then use the Markov chain Monte Carlo code \texttt{EMCEE} \cite{Foreman-Mackey:2012any}  to constrain the parameter space with the best-fit $\chi^2$-test, where the likelihood function $\mathcal{L}$ is estimated via $-2\ln\mathcal{L}=\chi^2=\chi^2_\mathrm{SNe}+\chi^2_\mathrm{BAO}+\chi^2_\mathrm{OHD(BC03/MS11)}$. For comparison, the $\Lambda$CDM model is also fitted to the same datasets with free parameters $\{\Omega_m, M_B, H_0, r_d\}$.

\section{Results}

The cosmological constraints from fitting the two different datasets, SNe+BAO+OHD(BC03) and SNe+BAO+OHD(MS11), to the $\Lambda$CDM and PAge models  are summarized in Table \ref{tab:constraint} and Fig. \ref{fig:MCMC}. For both $\Lambda$CDM and PAge models, the results from SNe+BAO+BC03 generally predict a lower $H_0$, a lower $M_B$, and a higher $r_d$ than those from SNe+BAO+MS11. The results from SNe+BAO+BC03 are closer to the usual constraints from CMB data, while the results from SNe+BAO+MS11 are  closer to the local direct measurements. This systematic shift might be caused by the different EPS models based on different empirical stellar libraries, for example, the MILES library \cite{Sanchez-Blazquez:2006mpa} for MS11 is slightly bluer (thus older age) than the STELIB library \cite{LeBorgne:2003te} for BC03. 
Nevertheless, this situation is similar to the inverse distance ladder constraint \cite{Arendse:2019hev} calibrated by SLTD from the H0LiCOW measurement \cite{Wong:2019kwg} on $H_0=73.3_{-1.8}^{+1.7}$ km/s/Mpc, which results in $r_d=(137\pm 3^\mathrm{stat.}\pm2^\mathrm{syst.})$ Mpc in tension with the CMB $r_d$ prior. This $r_d$ tension could be relaxed by calibrating the inverse distance ladder with the most recent SLTD measurement \cite{Birrer:2020tax} on $H_0=67.4_{-3.2}^{+4.1}$ km/s/Mpc from TDCOSMO+SLACS samples. Similar to the EPS-model dependence of the CC calibrator to the inverse distance ladder, the SLTD calibrator to the inverse distance ladder also admits an astrophysical dependence on the mass profile of lens galaxies. 

However, the key point is that, although the use of the different EPS models directly affects the cosmological constraints for the same model from different CC data, it affects identically both the PAge and $\Lambda$CDM models since CC data are independent of any cosmological models. The difference between the $\Lambda$CDM and PAge models fitted by the same datasets is negligibly small. The reduced minimal $\chi^2$  differs by 0.0016 (0.0014) between the $\Lambda$CDM and PAge models for SNe+BAO+BC03 (MS11). This could be made more quantitatively from the Bayesian information criterion (BIC) \cite{Schwarz:1978tpv} $\mathrm{BIC}=k\ln n-2\ln\mathcal{L}$, where $k$ is the number of the model parameters, and $n$ is the number of the data points. For SNe+BAO+BC03 (MS11), the BIC  difference of the PAge model with respect to the $\Lambda$CDM model is $\Delta\mathrm{BIC}=4.3(4.5)>2$. Therefore, there is positive evidence against the PAge model over the $\Lambda$CDM model.

Our PAge model is regarded here as a representative collection of various late-time models beyond $\Lambda$CDM. Different points in the $\eta-p_\mathrm{age}$ plane generally represent different models, and different models might also be degenerated at the same point in the $\eta-p_\mathrm{age}$ plane.  Matching the deceleration parameter of a specific model at different redshifts also results in different PAge representations. Therefore, our PAge model could cover a large number of late-time models. Furthermore, since both $r_d$ and $M_B$ are set as free fitting parameters in our data analysis, we also effectively cover those early-time models reducing to different values of $r_d$ and those astrophysical models with local calibrators to different values of $M_B$. Our final results then imply that there is a very little room for new physics beyond $\Lambda$CDM.

\section{Conclusions and discussions}

Despite the $\sim4\sigma$ tension in $H_0$ found between the global fitting result from the CMB data and that from the local distance ladder calibrated by Cepheids, the Hubble tension has been called into question for the potential unaccounted systematics \cite{Mortsell:2021nzg,Mortsell:2021tcx,Freedman:2021ahq}. Even if the Hubble tension turns out to be real, most of the early-time solutions run into tension with large-scale structure data \cite{Jedamzik:2020zmd}, while most of the late-time homogeneous solutions develop tension with the inverse distance ladder constraints \cite{Benevento:2020fev,Camarena:2021jlr,Efstathiou:2021ocp}, and the cosmic void as a late-time inhomogeneous solution \cite{GarciaBellido:2008nz,Keenan:2013mfa,Hoscheit:2018nfl} is also disfavored by the SNe data \cite{Wojtak:2013gda,Odderskov:2014hqa,Wu:2017fpr,Kenworthy:2019qwq,Lukovic:2019ryg,Cai:2020tpy}.

In this paper, we aim to generalize the late-time no-go argument with a global parametrizationfor the cosmic expansion history. Our final results slightly go against the representative PAge models over the $\Lambda$CDM model, which could be made tighter with inclusions of $f\sigma_8$ data \cite{Alestas:2021xes} reserved for future work.  No matter whether  the Hubble tension turns out to be real or not, our work could be regarded as a no-go guide for the Hubble solutions or a consistency test for the $\Lambda$CDM model.

If the Hubble tension persists to exist, then our work indicates that the Hubble solutions might come from some exotic modifications for our concordance Universe. For example, the early-time no-go argument \cite{Jedamzik:2020zmd} could be escaped from some EDE models \cite{Smith:2020rxx,Niedermann:2020qbw,Murgia:2020ryi,Allali:2021azp,Jiang:2021bab,Karwal:2021vpk} that could reduce the matter clumping. The late-time no-go arguments \cite{Benevento:2020fev,Camarena:2021jlr,Efstathiou:2021ocp} could be avoided by some inhomogeneous or anisotropic modifications \cite{Cai:2021wgv,Krishnan:2021dyb,Krishnan:2021jmh} for our local Universe or some modified gravity effects \cite{Desmond:2019ygn,Sakstein:2019qgn,SolaPeracaula:2019zsl,SolaPeracaula:2020vpg,Desmond:2020wep,Alestas:2020zol,Marra:2021fvf} for the magnitude-redshift relation. Some other hybrid models modifying both early-time and late-time Universe might still stand a chance in these $H_0$ Olympic-like games \cite{Schoneberg:2021qvd}.

If the Hubble tension disappears with improving calibration systematics, our work could be regarded as a consistency test for the $\Lambda$CDM model independent of early-Universe data and local $H_0$ measurements. The simple extensions of the $\Lambda$CDM model have already been tested with Refs.  \cite{Guo:2018ans,Vagnozzi:2019ezj} or without the CMB data \cite{Okamatsu:2021jil} from early-Universe observations and local $H_0$ measurements \cite{Dhawan:2017ywl}. Our work simply adds another layer of support for the $\Lambda$CDM model.

\begin{acknowledgments}
We thank Zhiqi Huang, Sunny Vagnozzi and Yuting Wang for the helpful discussions and correspondences.
This work is supported by the National Key Research and Development Program of China Grant No. 2020YFC2201501, No.2021YFC2203004, No.2021YFA0718304,
the National Natural Science Foundation of China Grants No. 11647601, No. 11690021, No. 11690022, No. 11821505, No. 11851302, No. 12047503, No. 11991052, No. 12075297, No. 12047558, and NO. 12105344,
the Strategic Priority Research Program of the Chinese Academy of Sciences (CAS) Grant No. XDB23030100, No. XDA15020701, 
the Key Research Program of the CAS Grant No. XDPB15, 
the Key Research Program of Frontier Sciences of CAS, 
the China Postdoctoral Science Foundation Grant No. 2021M693238, 
the Special Research Assistant Funding Project of CAS,
and the Science Research Grants from the China Manned Space Project with No. CMS-CSST-2021-B01.
\end{acknowledgments}

\appendix

\section{Taylor expansions}\label{app:Taylor}

For self-contained, we list below the Taylor expansions \cite{Cattoen:2007sk}  in redshift $z$ \cite{Zhang:2016urt} and $y$-redshift $y\equiv1-a=z/(1+z)$ \cite{Capozziello:2011tj}  for dimensionless Hubble expansion rate $E=H/H_0$ and dimensionless luminosity distance $d_L=D_L/(c/H_0)$, respectively.

\begin{align}
E(z)&=1+(1+q_0)z+\frac12(-q_0^2+j_0)z^2+\frac16(3q_0^2+3q_0^3\nonumber\\
&-4q_0j_0-3j_0-s_0)z^3+\frac{1}{24}(-12q_0^2-24q_0^3-15q_0^4\nonumber\\
&+32q_0j_0+25q_0^2j_0+7q_0s_0+12j_0-4j_0^2+8s_0+l_0)z^4\nonumber\\
&+\mathcal{O}(z^5)
\end{align}

\begin{align}
d_L(z)&=z+\frac12(1-q_0)z^2+\frac16(-1+q_0+3q_0^2-j_0)z^3\nonumber\\
&+\frac{1}{24}(2-2q_0-15q_0^2-15q_0^3+5j_0+10q_0j_0+s_0)z^4\nonumber\\
&+\frac{1}{120}(-6+6q_0+81q_0^2+165q_0^3+105q_0^4+10j_0^2\nonumber\\
&-27j_0-110q_0j_0-105q_0^2j_0-15q_0s_0-11s_0-l_0)z^5\nonumber\\
&+\mathcal{O}(z^6)
\end{align}

\begin{align}
E(y)&=1+(1+q_0)y+(1+q_0-\frac12q_0^2+\frac12j_0)y^2\nonumber\\
&+\frac16(6+6q_0-3q_0^2+3q_0^3+3j_0-4q_0j_0-s_0)y^3\nonumber\\
&+\frac{1}{24}(24+24q_0-12q_0^2+12q_0^3-15q_0^4+12j_0\nonumber\\
&-4j_0^2-16q_0j_0+25q_0^2j_0-4s_0+7q_0s_0+l_0)y^4\nonumber\\
&+\mathcal{O}(y^5)
\end{align}

\begin{align}
d_L(y)&=y+\frac32(1-q_0)y^2+\frac16(11-5q_0+3q_0^2-j_0)y^3\nonumber\\
&+\frac{1}{24}(50-26q_0+21q_0^2-15q_0^3-7j_0+10q_0j_0+s_0)y^4\nonumber\\
&+\frac{1}{120}(274-154q_0+141q_0^2-135q_0^3+105q_0^4+10j_0^2\nonumber\\
&-47j_0+90q_0j_0-105q_0^2j_0-15q_0s_0+9s_0-l_0)y^5\nonumber\\
&+\mathcal{O}(y^6)
\end{align}

Here the Hubble, deceleration, jerk, snap, and lerk parameters are defined as
\begin{align}
H(t)&\equiv+\frac{1}{a}\frac{\mathrm{d}a}{\mathrm{d}t},\\
q(t)&\equiv-\frac{1}{a}\frac{\mathrm{d}^2a}{\mathrm{d}t^2}\left(\frac{1}{a}\frac{\mathrm{d}a}{\mathrm{d}t}\right)^{-2},\\
j(t)&\equiv+\frac{1}{a}\frac{\mathrm{d}^3a}{\mathrm{d}t^3}\left(\frac{1}{a}\frac{\mathrm{d}a}{\mathrm{d}t}\right)^{-3},\\
s(t)&\equiv+\frac{1}{a}\frac{\mathrm{d}^4a}{\mathrm{d}t^4}\left(\frac{1}{a}\frac{\mathrm{d}a}{\mathrm{d}t}\right)^{-4},\\
l(t)&\equiv+\frac{1}{a}\frac{\mathrm{d}^5a}{\mathrm{d}t^5}\left(\frac{1}{a}\frac{\mathrm{d}a}{\mathrm{d}t}\right)^{-5},
\end{align}
respectively. For $\Lambda$CDM with late-time parametrization $H(t)=H_0\sqrt{\Omega_ma(t)^{-3}+(1-\Omega_m)}$, the current value for the deceleration, jerk, snap, and lerk parameters read
\begin{align}
q_0&=-1+\frac32\Omega_m,\\
j_0&=1,\\
s_0&=1-\frac92\Omega_m,\\
l_0&=1+\frac32\Omega_m(2+9\Omega_m),
\end{align}
respectively.

\section{Model matching}\label{app:model}

\begin{figure}
\centering
\includegraphics[width=0.48\textwidth]{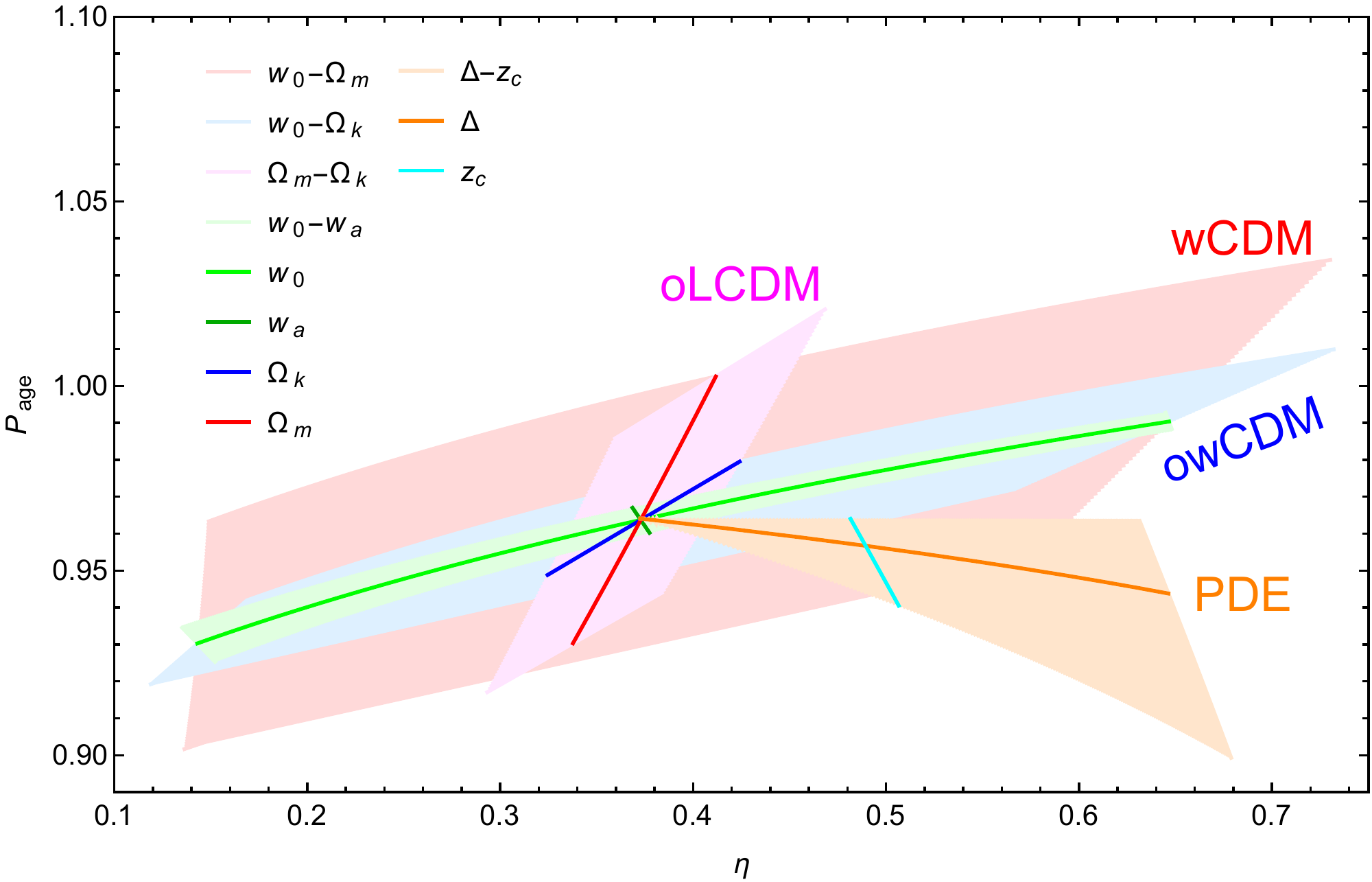}\\
\caption{PAge representations of the $ow_\mathrm{CPL}\mathrm{CDM}$ model and PDE model in the $\eta-p_\mathrm{age}$ space. Changing one or two of the model parameters within $0.26\leq\Omega_m\leq0.34$, $-0.05\leq\Omega_k\leq0.05$, $-1.2\leq w_0\leq-0.8$, $-0.1\leq w_a\leq0.1$ and $0\leq\Delta\leq1$, $0\leq0.001<z_c<0.5$ are shown in color lines and shaded regions as indicated. The fiducial reference cosmology is chosen as  $H_0=70\,\mathrm{km/s/Mpc}$,  $\Omega_m=0.3$, $\Omega_k=0$, $w_0=-1$, $w_a=0$ for the $ow_\mathrm{CPL}\mathrm{CDM}$ model and $\Delta=0.3$, $z_c=0.1$, $\beta=2$ for the PDE model.}\label{fig:PAgeModel}
\end{figure}

As the simplest example, $\Lambda$CDM could be extended with a nonzero $\Omega_k$, and a dynamical dark energy with Chevallier-Polarski-Linder (CPL) parametrization$w(a)=w_0+w_a(1-a)$ \cite{Chevallier:2000qy,Linder:2002et}.
For this $ow_\mathrm{CPL}\mathrm{CDM}$ model with
\begin{align}
E(a)^2&=\Omega_ma^{-3}+\Omega_ka^{-2}\\
&+(1-\Omega_m-\Omega_k)a^{-3(1+w_0+w_a)}e^{-3w_a(1-a)},\nonumber
\end{align}
the deceleration parameter in \eqref{eq:eta} is therefore expressed by the model parameters as
\begin{align}
q_0=\frac12(1-\Omega_k)+\frac32(1-\Omega_k-\Omega_m)w_0,
\end{align}
and so does the current age of our Universe $t_0$, both of which could be mapped to the $\eta-p_\mathrm{age}$ space as shown in Fig. \ref{fig:PAgeModel}. As shown in Table 1 of \cite{Luo:2020ufj}, the relative error of $|\Delta D_A|/D_A$ for  PAge representation of this $ow_\mathrm{CPL}\mathrm{CDM}$ model can be less than $1\%$ over $0<z<2.5$. 

\begin{figure}
\centering
\includegraphics[width=0.48\textwidth]{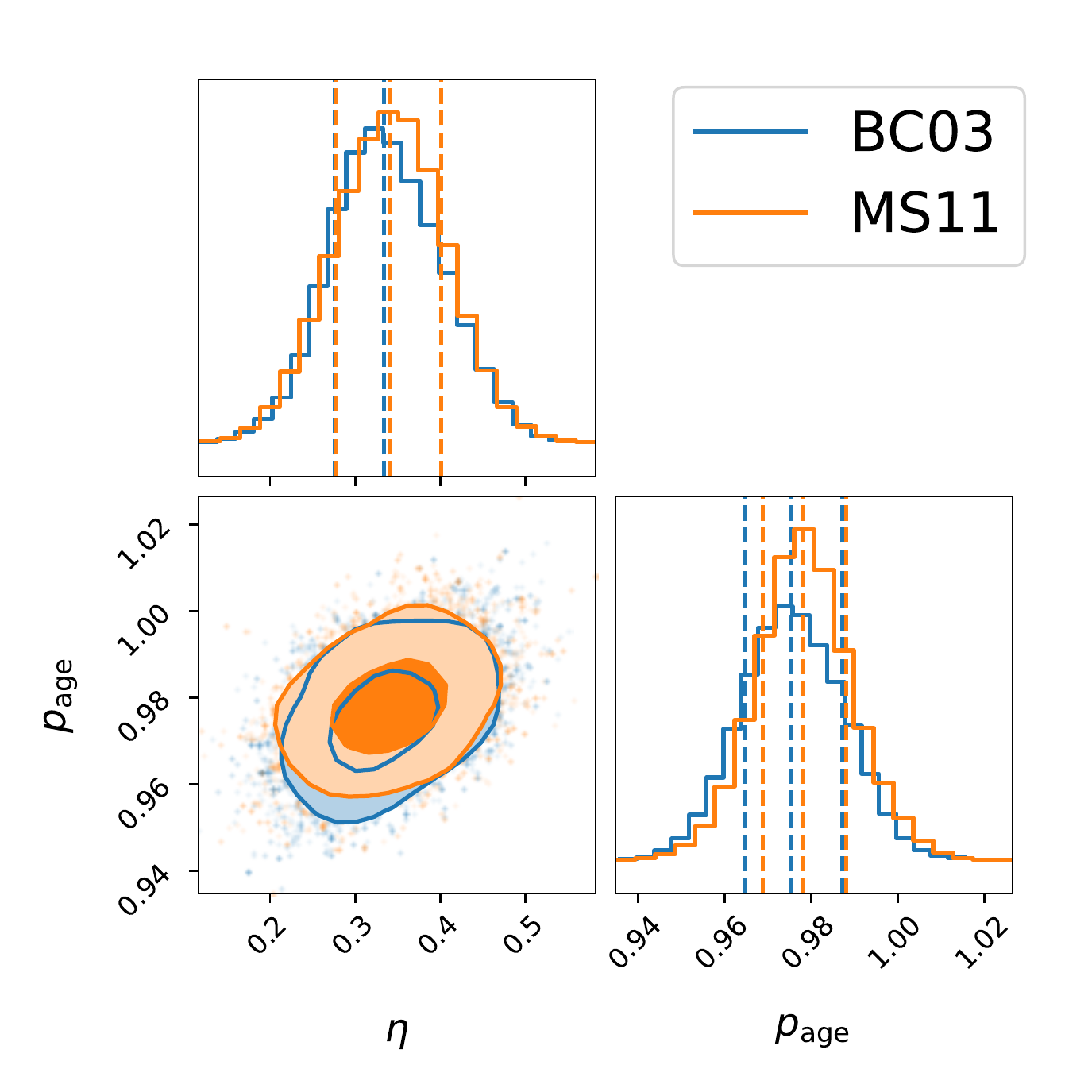}\\
\caption{$1\sigma$ and $2\sigma$ constraints on the PAge parameters extracted from the full constraints of fitting the datasets SNe+BAO+OHD(BC03) (blue) and SNe+BAO+OHD(MS11) (orange) to a general PAge model with free parameters $\{\eta, p_\mathrm{age}, M_B, H_0, r_d\}$. }\label{fig:PAge}
\end{figure}

A second example is the phantom-like dark energy (PDE) model \cite{Efstathiou:2021ocp} with the dimensionless Hubble parameter defined by 
\begin{align}
E(z)^2=\Omega_m(1+z)^3+(1-\Omega_m)\left(1+\Delta e^{-\left(\frac{z}{z_c}\right)^\beta}\right),
\end{align}
where $z_c$ is a phantom-like transition redshift and $\Delta$ value characterizes the strength of the phantom-like behavior. The deceleration parameter could be calculated as
\begin{align}
q_0=-1+\frac32\frac{\Omega_m}{1+(1-\Omega_m)\Delta}.
\end{align}
Now the PAge parameters $\{\eta,p_\mathrm{age}\}=\{1-\frac32p_\mathrm{age}^2(1+q_0), H_0t_0\}$ could be directly expressed in terms of the model parameters $\{\Delta, z_c, \beta\}$, which could be presented in the PAge parameter space as shown in Fig. \ref{fig:PAgeModel}. The relative error of $|\Delta H|/H$ for the PAge representation of PDE is less than 5\% as one can explicitly check for the redshift range $0.1\lesssim z\lesssim 10^3$. As shown in Fig. \ref{fig:PAge}, the parameter region of this PDE model in PAge parameter space is largely outside the $2\sigma$ contour  from fitting the datasets SNe+BAO+OHD(BC03) and SNe+BAO+OHD(MS11) to a general PAge model with free parameters $\{\eta, p_\mathrm{age}, M_B, H_0, r_d\}$. This confirms the previous study \cite{Efstathiou:2021ocp} from a different point of view.

\section{Data}\label{app:Data}

For convenience, we also list below the full data points of BAO and OHD we used in the data analysis. 

The state-of-art datasets for BAO include
\begin{itemize}
\item Six-degree Field Galaxy Survey (6dFGS) \cite{Beutler:2011hx},
\item Sloan Digital Sky Survey Data Release 7 Main Galaxy Sample (SDSS DR7 MGS) \cite{Ross:2014qpa},
\item SDSS-III Baryon Oscillation Spectroscopic Survey DR12 (SDSS-III BOSS DR12) \cite{BOSS:2016zkm} (Note that this tomographic constraints are more sensitive to the dynamical dark energy than the consensus 3-bin BAO and RSD measurements at three effective redshifts in \cite{Alam:2016hwk}),
\item SDSS-IV eBOSS DR14 quasar sample (SDSS-IV eBOSS DR14 QSO) \cite{Ata:2017dya}, 
\item SDSS-IV eBOSS DR14 Ly$\alpha$ \cite{deSainteAgathe:2019voe}, 
\item SDSS-IV eBOSS DR14 QSO-Ly$\alpha$ \cite{Blomqvist:2019rah}, 
\item SDSS-IV extended BOSS DR16 Luminous Red Galaxies (SDSS-IV eBOSS DR14 LRG) \cite{Bautista:2020ahg,Gil-Marin:2020bct} (Note that this result is actually inferred from the DR16 CMASS + eBOSS LRG galaxies),
\item SDSS-IV eBOSS DR16 emission line galaxies (SDSS-IV eBOSS DR16 ELG) \cite{deMattia:2020fkb,Tamone:2020qrl}, 
\item SDSS-IV eBOSS DR16 QSO \cite{Neveux:2020voa,Hou:2020rse}, 
\item SDSS-IV eBOSS DR16 Ly$\alpha$ \cite{duMasdesBourboux:2020pck},
\item Dark Energy Survey  Year 3 (DES Y3) \cite{DES:2021esc}. 
\end{itemize}

Note that the OHD points at $z=1.363$ and $z=1.965$ from \cite{Moresco:2015cya} adopt both EPS models of BC03 and MS11, which are not included in our presented results. Nevertheless, we have checked that the naive inclusion of these two OHD points in both datasets has little impact on our results and conclusion.

\begin{table}
\caption{BAO data}\label{tab:BAO}
	\centering
	\begin{tabular}{cccr}
		\hline
		\hline
		$z_\mathrm{eff}$ & Measurement & Constraint & References \\
		\hline
		& & &  6dFGS \\
		$0.106$ & $r_d/D_V$ & $0.336 \pm 0.015$ & \cite{Beutler:2011hx}\\
		\hline
		& & &  SDSS DR7 MGS \\
		$0.15$ & $D_V/r_d$ & $4.47 \pm 0.17$ &  \cite{Ross:2014qpa} \\
		\hline
		 & & & SDSS BOSS DR12 \\
		$0.31$ & $D_A/r_d$ & $6.29 \pm 0.14$ & \cite{BOSS:2016zkm} \\
		$0.36$ & $D_A/r_d$ & $7.09 \pm 0.16$ & \cite{BOSS:2016zkm}\\
		$0.40$ & $D_A/r_d$ & $7.70 \pm 0.16$ & \cite{BOSS:2016zkm}\\
		$0.44$ & $D_A/r_d$ & $8.20 \pm 0.13$ & \cite{BOSS:2016zkm}\\
		$0.48$ & $D_A/r_d$ & $8.64 \pm 0.11$ & \cite{BOSS:2016zkm}\\
		$0.52$ & $D_A/r_d$ & $8.90 \pm 0.12$ & \cite{BOSS:2016zkm}\\
		$0.56$ & $D_A/r_d$ & $9.16 \pm 0.14$ & \cite{BOSS:2016zkm}\\
		$0.59$ & $D_A/r_d$ & $9.45 \pm 0.17$ & \cite{BOSS:2016zkm}\\
		$0.64$ & $D_A/r_d$ & $9.62 \pm 0.22$ & \cite{BOSS:2016zkm}\\
		$0.31$ & $H*r_d$ & $11550 \pm 700$ & \cite{BOSS:2016zkm}\\
		$0.36$ & $H*r_d$ & $11810 \pm 500$ & \cite{BOSS:2016zkm}\\
		$0.40$ & $H*r_d$ & $12120 \pm 300$ & \cite{BOSS:2016zkm}\\
		$0.44$ & $H*r_d$ & $12530 \pm 270$ & \cite{BOSS:2016zkm}\\
		$0.48$ & $H*r_d$ & $12970 \pm 300$ & \cite{BOSS:2016zkm}\\
		$0.52$ & $H*r_d$ & $13940 \pm 390$ & \cite{BOSS:2016zkm}\\
		$0.56$ & $H*r_d$ & $13790 \pm 340$ & \cite{BOSS:2016zkm}\\
		$0.59$ & $H*r_d$ & $14550 \pm 470$ & \cite{BOSS:2016zkm}\\
		$0.64$ & $H*r_d$ & $14600 \pm 440$ & \cite{BOSS:2016zkm}\\
		\hline
		& & & eBOSS DR14 QSO \\
		$1.52$ & $D_V/r_d$ & $26.00 \pm 0.99$ &  \cite{Ata:2017dya}\\
		\hline
		& & & eBOSS DR14 Ly$\alpha$ \\
		$2.34$ & $D_H/r_d$ & $8.86 \pm 0.29$ &  \cite{deSainteAgathe:2019voe}\\
		$2.34$ & $D_M/r_d$ & $37.41 \pm 1.86$ & \cite{deSainteAgathe:2019voe}\\
		\hline
		& & & eBOSS DR14 QSO-Ly$\alpha$ \\
		$2.35$ & $D_H/r_d$ & $9.20 \pm 0.36$ & \cite{Blomqvist:2019rah} \\
		$2.35$ & $D_M/r_d$ & $36.3 \pm 1.8$ & \cite{Blomqvist:2019rah} \\
		\hline
		 & & & eBOSS DR16 LRG \\
		$0.698$ & $D_H/r_d$ & $19.77 \pm 0.47$ & \cite{Bautista:2020ahg,Gil-Marin:2020bct}\\
		$0.698$ & $D_M/r_d$ & $17.65 \pm 0.30$ & \cite{,Bautista:2020ahg,Gil-Marin:2020bct}\\
		\hline
		& & & eBOSS DR16 ELG \\
		$0.845$ & $D_V/r_d$ & $18.33_{-0.62}^{+0.57}$ & \cite{deMattia:2020fkb}\\
		$0.85$ & $D_H/r_d$ & $19.6 _{-2.1}^{+2.2}$ & \cite{deMattia:2020fkb,Tamone:2020qrl} \\
		$0.85$ & $D_M/r_d$ & $19.5 \pm 1.0$ & \cite{deMattia:2020fkb,Tamone:2020qrl} \\
		\hline
		& & & eBOSS DR16 QSO \\
		$1.48$ & $D_H/r_d$ & $13.23 \pm 0.47$ & \cite{Neveux:2020voa,Hou:2020rse} \\
		$1.48$ & $D_M/r_d$ & $30.21 \pm 0.79$ & \cite{Neveux:2020voa,Hou:2020rse} \\
		\hline
		& & & eBOSS DR16 Ly$\alpha$ \\
		$2.33$ & $D_H/r_d$ & $8.99 \pm 0.19$ & \cite{duMasdesBourboux:2020pck}\\
		$2.33$ & $D_M/r_d$ & $37.5 \pm 1.1$ & \cite{duMasdesBourboux:2020pck}\\
		\hline
		& & & DES Y3 \\
		$0.835$ & $D_M/r_d$ & $18.92 \pm 0.51$ & \cite{DES:2021esc} \\
		\hline
		\hline
	\end{tabular}
\end{table}

\begin{table}
\caption{OHD with EPS model from BC03}\label{tab:OHDBC03}
	\begin{tabular}{ccc}
		\hline
		\hline
		$z$ & $H(z)\,\mathrm{km/s/Mpc}$  & Reference \\
		\hline
		0.3802 & $83.0 \pm 13.5$ & \cite{Moresco:2016mzx} \\
		0.4004 & $77.0 \pm 10.2$ & \cite{Moresco:2016mzx} \\
		0.4247 & $87.1 \pm 11.2$ & \cite{Moresco:2016mzx} \\
		0.4497 & $92.8 \pm 12.9$ & \cite{Moresco:2016mzx} \\
		0.4783 & $80.9 \pm 9$ & \cite{Moresco:2016mzx} \\
		0.4293 & $85.7 \pm 5.2$ & \cite{Moresco:2016mzx} \\
		0.1791 & $75 \pm 4$ & \cite{Moresco:2012jh} \\
		0.1993 & $75 \pm 5$ & \cite{Moresco:2012jh} \\
		0.3519 & $83 \pm 14$ & \cite{Moresco:2012jh} \\
		0.5929 & $104 \pm 13$ & \cite{Moresco:2012jh} \\
		0.6797 & $92 \pm 8$ & \cite{Moresco:2012jh} \\
		0.7812 & $105 \pm 12$ & \cite{Moresco:2012jh} \\
		0.8754 & $125 \pm 17$ & \cite{Moresco:2012jh} \\
		1.037 & $154 \pm 20$ & \cite{Moresco:2012jh} \\
		0.07 & $69.0 \pm 19.6$ & \cite{Zhang:2012mp} \\
		0.12 & $68.6 \pm 26.2$ & \cite{Zhang:2012mp} \\
		0.20 & $72.9 \pm 29.6$ & \cite{Zhang:2012mp} \\
		0.28 & $88.8 \pm 36.6$ & \cite{Zhang:2012mp} \\
		0.47 & $89 \pm 23$ & \cite{Ratsimbazafy:2017vga}\\
		0.1 & $69 \pm 12$ & \cite{Stern:2009ep}  \\
		0.17 & $83 \pm 8$ & \cite{Stern:2009ep} \\
		0.27 & $77 \pm 14$ & \cite{Stern:2009ep} \\
		0.4 & $95 \pm 17$ & \cite{Stern:2009ep}  \\
		0.48 & $97 \pm 62$ & \cite{Stern:2009ep}  \\
		0.88 & $90 \pm 40$ & \cite{Stern:2009ep} \\
		0.9 & $117 \pm 23$ & \cite{Stern:2009ep}  \\
		1.3 & $168 \pm 17$ & \cite{Stern:2009ep} \\
		1.43 & $177 \pm 18$ & \cite{Stern:2009ep} \\
		1.53 & $140 \pm 14$ & \cite{Stern:2009ep} \\
		1.75 & $202 \pm 40$ & \cite{Stern:2009ep} \\
		\hline
		1.363 & $160\pm33.6$ & \cite{Moresco:2015cya}\\
		1.965 & $186.5\pm50.4$ & \cite{Moresco:2015cya}\\
		\hline
		\hline
	\end{tabular}
\end{table}

\begin{table}
\caption{OHD with EPS model from MS11}\label{tab:OHDMS11}
	\begin{tabular}{ccc}
		\hline
		\hline
		$z$ & $H(z)\,\mathrm{km/s/Mpc}$  & Reference \\
		\hline
		0.3802 & $89.3 \pm 14.1$ & \cite{Moresco:2016mzx}\\
		0.4004 & $82.8 \pm 10.6$ & \cite{Moresco:2016mzx} \\
		0.4497 & $99.7 \pm 13.4$ & \cite{Moresco:2016mzx}\\
		0.4247 & $93.7 \pm 11.7$ & \cite{Moresco:2016mzx} \\
		0.4783 & $86.6 \pm 8.7$ & \cite{Moresco:2016mzx}\\
		0.4293 & $91.8 \pm 5.3$ & \cite{Moresco:2016mzx}\\
		0.1791 & $81 \pm 5$ & \cite{Moresco:2012jh} \\
		0.1993 & $81 \pm 6$ & \cite{Moresco:2012jh} \\
		0.3519 & $88 \pm 16$ & \cite{Moresco:2012jh} \\
		0.5929 & $110 \pm 15$ & \cite{Moresco:2012jh} \\
		0.6797 & $98 \pm 10$ & \cite{Moresco:2012jh} \\
		0.7812 & $88 \pm 11$ & \cite{Moresco:2012jh} \\
		0.8754 & $124 \pm 17$ & \cite{Moresco:2012jh} \\
		1.037 & $113 \pm 15$ & \cite{Moresco:2012jh} \\
		\hline
		1.363 & $160\pm33.6$ & \cite{Moresco:2015cya}\\
		1.965 & $186.5\pm50.4$ & \cite{Moresco:2015cya}\\
		\hline
		\hline
	\end{tabular}
\end{table}


\bibliographystyle{utphys}
\bibliography{ref}

\end{document}